# Relationship between the DC Bias and Debye Length in a Complex Plasma

Jie Kong, Jorge C. Reyes, James Creel, and Truell Hyde, *Member, IEEE*

*Abstract*— **The levitation height of a dust particle layer within a RF discharge plasma sheath is known to be related to the DC bias, the background pressure, and the Debye length. In this paper, a new experimental technique for measurement of the Debye length is introduced. This technique is based on the relationship between an externally applied DC bias and the particle levitation height and shows that under appropriate conditions, the addition of an externally applied DC bias provides a mechanism for evaluation of the Debye length. When compared with existing techniques, this new method appears to be simpler to implement in some cases.**

## I. INTRODUCTION

The Debye shielding length is one of the most important system parameters employed in the study of dusty plasmas due to its association with the shielding length scale for external electrostatic fields. This is due in part to the fact that the Yukawa potential, plasma frequency, plasma density, and coupling constant can all be directly related to the Debye length [1]. The definition of Debye length is,

$$\lambda_D^{-2} = \lambda_{Di}^{-2} + \lambda_{De}^{-2} \text{ ,} \qquad (1a)$$

$$\lambda_{Di,e} = \sqrt{\frac{\varepsilon_0 K_B T_{i,e}}{e^2 n_{i,e}}} \text{ ,}$$

(1b)

where $T_{i,e}$ are the ion and electron temperatures respectively, $n_{i,e}$ are the ion and electron densities, $K_B$ is the Boltzmann constant, $\varepsilon_0$ is the free space dielectric constant, and $e$ is the elementary charge. The definition of Debye length in this form provides a straight forward experimental method for its determination by a simple measurement of the plasma density. Employing any standard technique, such as the well known Langmuir probe method [2]-[5], the Debye length of the plasma can be derived from the measured electron and ion densities, as well as the corresponding electron temperature. Unfortunately, the Langmuir probe method is perturbative by nature, complicating an accurate determination of $\lambda_D$. Several new experimental techniques for the determination of the dusty plasma Debye shielding length have recently been developed. One of the technique is using dust lattice waves [6], [7]. In this case, the vertical resonant frequency of the dust particles is defined as $\omega_0 = \sqrt{Q_D e n_i / \varepsilon_0 M_D}$ , where $Q_D$ and $M_D$ are the dust particle's charge and mass respectively, $n_i$ is the ion density, and $e$ is the elementary charge. The resonant frequency $\omega_0$ can be experimentally measured once excited employing perturbation of the signal driving the lower electrode or via laser excitation. However, as can be seen in the above relationship, $\omega_0$ is a function of both the dust particle charge, $Q_D$ and ion density, $n_i$. As such, to determine the Debye length from $n_i$, the particle charge $Q_D$ must first be determined employing an independent experimental method, again complicating the determination of $\lambda_D$ [8]. Other techniques for the determination of Debye length based on the phonon dispersion relation [9]-[11] have also been developed. One of the primary parameters required in such wave dispersion relations is the shielding parameter $\kappa = a/\lambda_D$ , which can be derived from the experimental dispersion relations.

In this paper, a new experimental technique for determining the Debye length within a complex plasma is introduced. This method provides a mechanism for calculation of the Debye length employing the measured change in dust particle levitation height caused by a change in the external DC bias.

## II. THEORETICAL BACKGROUND

It is well known that the fast moving electrons near a confinement wall (such as the lower electrode in a GEC rf reference cell) build up a negative potential on the wall [12]. The resulting sheath is generally confined by the Debye shielding length to a layer of only a few Debye lengths. Therefore the thickness of this sheath, $d_{sh}$ , can be

Manuscript received August 4, 2006. This work was supported in part by National Science Foundation Grant PHY-0353558.

Jie Kong and Truell Hyde are with the Center for Astrophysics, Space Physics, and Engineering Research, Baylor University, Waco, TX 76798 USA (phone: 254-710-3763; e-mail: J_Kong@baylor.edu, Truell_Hyde@baylor.edu).

James Creel was a 2006 REU fellow with Baylor University, Waco, TX 76798 USA.

expressed as a function of the powered electrode potential (resulting in this case from both the self-induced and external DC bias), $\phi_w$, and the Debye length, $\lambda_D$, as given by

$$d_{sh} = d_{sh}(\phi_w, \lambda_D)\,. \tag{2}$$

Any small change in the DC bias, from $\phi_{w1}$ to $\phi_{w2}$, will introduce a subsequent change in sheath thickness

$$\Delta d_{sh} = \Delta d_{sh}(\lambda_D)\,, \tag{3a}$$

$$\Delta d_{sh} = d_{sh2} - d_{sh1}$$

$$= d_{sh}(\phi_{w2}, \lambda_D) - d_{sh}(\phi_{w1}, \lambda_D)\,, \tag{3b}$$

where it is assumed the difference in DC bias is small enough that it will not cause a change in either the Debye length or other system parameters. Under this assumption, Equation 3 can now be used to determine the Debye length experimentally.

In the above equation, the sheath thickness is determined employing an analytical expression derived from the collisional sheath model [13]. The sheath model assumes to be an unmagnetized, charge-neutral plasma in contact with a planar wall. In the plasma both the density of electrons $n_e$ and the density of (positive) ions $n_i$ are equal to the plasma density $n_0$. The potential within the sheath is $\phi$, and the wall is held at a negative potential $\phi_w$. Under these conditions, a sheath will form separating the plasma from the wall where the sheath is assumed to be source-free. The governing equations for such a situation are based on a two-fluid model where the electrons obey the Boltzmann relation, and the cold ions obey the source-free, steady state equation of continuity. For this situation, Poisson's equation relates the electron and ion densities to the self-consistent potential and the ion-neutral collision cross section can be modeled using a power law dependence on ion energy. An analytic relation for the sheath thickness can then be derived by solving Poisson's equation:

:

$$d_{sh} = \lambda_D$$

$$\cdot\left[\frac{(5+2\gamma)^{3+\gamma}\cdot(2+\gamma)^{2+\gamma}}{(3+\gamma)^{5+2\gamma}}\cdot\frac{\eta_w^{3+\gamma}}{\alpha\cdot u_0^{3+\gamma}}\right]^{1/(5+2\gamma)}, \tag{4}$$

where $\alpha$ is the collision parameter defined as $\alpha = \lambda_D / \lambda_{mfp}$ with $\lambda_{mfp}$ being the ion mean free path, $\lambda_{mfp} = 1/(n_n \sigma_s)$, $n_n$ is the neutral gas density, $\sigma_s$ is the ion-neutral collision cross section at the ion acoustic speed (for an Argon plasma at the Bohm velocity, $\sigma_s = 8\times10^{-19} m^2$ [14]), $\gamma$ is a dimensionless parameter ranging from 0 to -1 (for a constant mean free path $\gamma = 0$, and for constant ion mobility $\gamma = -1$), $u_0 = V_0 / C_s$ is the dimensionless ion speed at the sheath entry, $C_s = \sqrt{K_B T_e / m_i}$ is the ion acoustic velocity and $V_0$ is the ion speed at the sheath entry. A value of $V_0 = 1.1C_s$ was used in this calculation. $\eta_w = -e\phi_w / K_B T_e$. $\phi_w$ is the electrode potential and $T_e$ is the electron temperature. Figure 1 shows a plot of $\Delta d_{sh}$ as a function of $\lambda_D$, calculated using Equations 3 and 4.

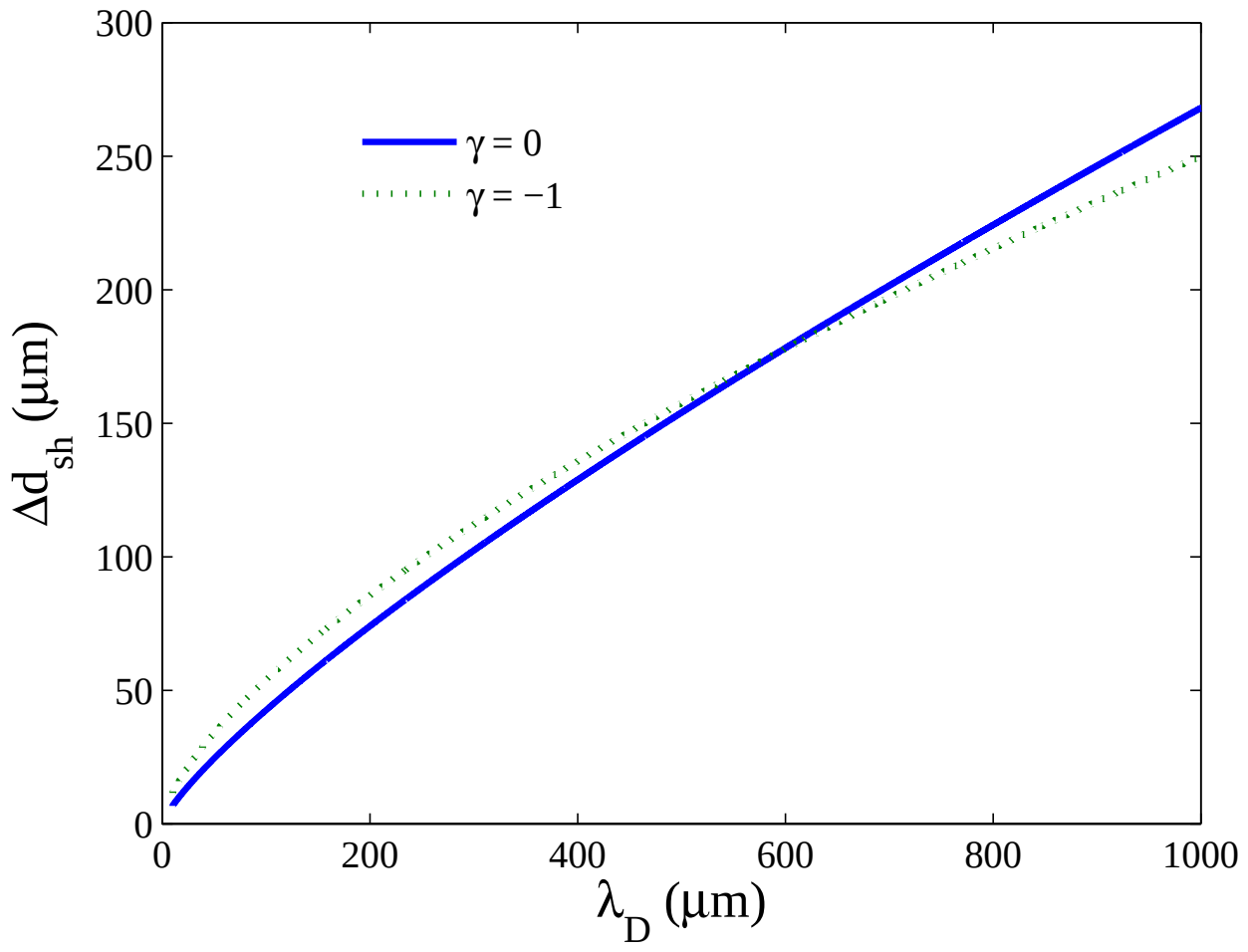

Fig. 1. Calculated difference in sheath thickness as a function of the Debye length. The Argon gas pressure is 26.7 Pa (200 mTorr) and the bias potentials are $\phi_{w1} = -48.8V$, and $\phi_{w2} = -56.2V$.

For this case, the Argon gas pressure was 26.7 Pa (200 mTorr), the RF power was 5W, the DC bias ranged from $\phi_{w1} = -48.8V$ to $\phi_{w2} = -56.2V$, and the electron temperature was $K_B T_e = 5.10eV$. The two curves shown are for $\gamma = 0$ and $\gamma = -1$ as indicated in the figure. As can be seen, the monotonic nature of $\Delta d_{sh}$ as a function of $\lambda_D$ provides a unique value of Debye length for a given change in sheath thickness. As such, if the change in sheath thickness created by a small difference in DC bias is measured experimentally, the system Debye length can then be derived from the matching point using the relationship shown in Figure 1. As an example, under the above mentioned experimental settings, if the measured change in sheath thickness is $100\mu m$, from Figure 1 the corresponding Debye length will be between $250\mu m$ for $\gamma = -1$, and $290\mu m$ for $\gamma = 0$. For simplicity, the median value $\gamma = -0.5$ is used in the following calculations to determine $\lambda_D$, with the difference between the two extreme cases, $\gamma = 0$, and $\gamma = -1$, included as part of the error analysis.

Experimentally, any change in DC bias causes a change in the potential distribution within the sheath, and subsequently, the Debye length is not constant in Equation 3.

Therefore, the Debye length as derived above will have an error term,

$$\lambda_D = \lambda_{D0} + \delta\lambda_D(\Delta\phi_w) \quad (5a)$$

$$\Delta\phi_w = \phi_{w2} - \phi_{w1}, \quad (5b)$$

where $\lambda_{D0}$ is the Debye length value at $\Delta\phi_w = 0$, and $\delta\lambda_D \to 0$ as $\Delta\phi_w \to 0$. This error term includes effects such as the change in dust charge (due to the addition of dust particles to the sheath), the change in potential distribution, experimental error etc. From Equation 3, $\Delta d_{sh} \to 0$ as $\Delta\phi_w \to 0$, where experimentally the measurement error increases as $\Delta d_{sh} \to 0$, since it is difficult to measure $\Delta d_{sh}$ at small $\Delta\phi_w$. Once the relationship of $\lambda_D$ as a function of $\Delta\phi_w$ is determined experimentally, $\lambda_{D0}$ can be derived from the relationship $\lambda_D = \lambda_D(\Delta\phi_w)$ by extrapolating $\lambda_D(\Delta\phi_w = 0)$.

## III. EXPERIMENT

As a test of the method given above, an experiment was conducted employing the CASPER GEC rf reference cell. A radio-frequency, capacitively coupled discharge was formed between two parallel-plate electrodes, 8 cm in diameter and separated by 3 cm, with the bottom electrode water-cooled. The lower electrode is powered by a radio-frequency signal generator, while the upper electrode is grounded as is the chamber. The signal generator is coupled to the electrode through an impedance matching network and a variable capacitor attenuator network. The plasma discharge apparatus is described in greater detail in [15], [16].

To measure the sheath thickness accurately is difficult, particularly when dust particles are present. However, the change in the dust particle layer levitation height can usually be easily measured. Since both the sheath thickness and the dust layer levitation height change with the DC bias, we make the assumption they are equal to first order, $\Delta d_{sh} = \Delta Z$, when the change in DC bias is small. ($\Delta Z$ is the change in dust layer height as shown in Figure 2.)

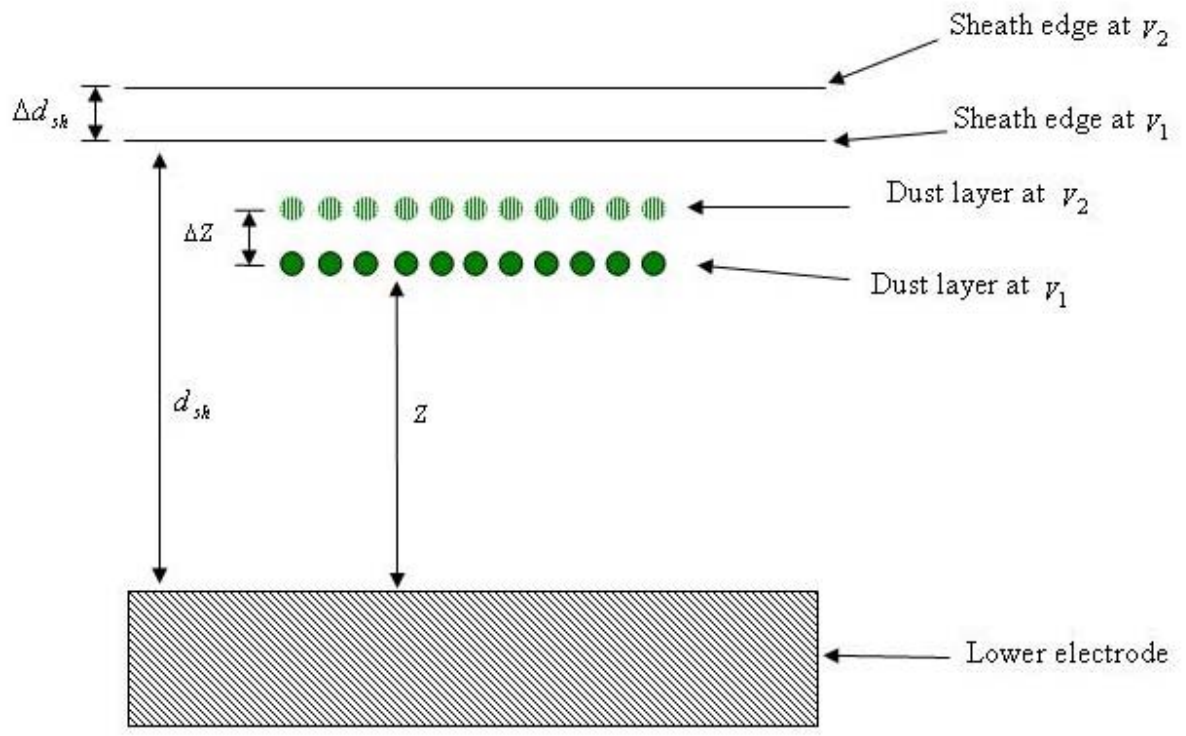

Fig. 2. The change in bias on the lower electrode is given as $\Delta V = V_2 - V_1$, the sheath thickness is $d_{sh}$, and the dust particle layer is considered to be at a height $Z$ above the lower electrode at $V_1$. A small change in bias, $\Delta V$, induces a corresponding change in sheath thickness $\Delta d_{sh}$, and in the height of the dust particle layer, $\Delta Z$.

To verify this assumption, the change in dust layer height and sheath thickness as a function of the DC bias was measured experimentally using a side camera without filters in order to allow capture of both the reflected laser light from the particles and the plasma glow emission (see Fig. 3). The dust particles shown are MF (melamine formaldehyde) and are 8.9 $\mu m$ in diameter. A horizontal laser sheet was used to illuminate the dust crystal for view by a top mounted camera, while a vertical laser sheet illuminated the dust levitation level for the side mounted camera. Figure 3 shows both the position of the dust particle layer and the bottom edge of the plasma. As can be seen in Fig. 3, the dust particle position can be easily determined, whereas the sheath edge can not be clearly defined.

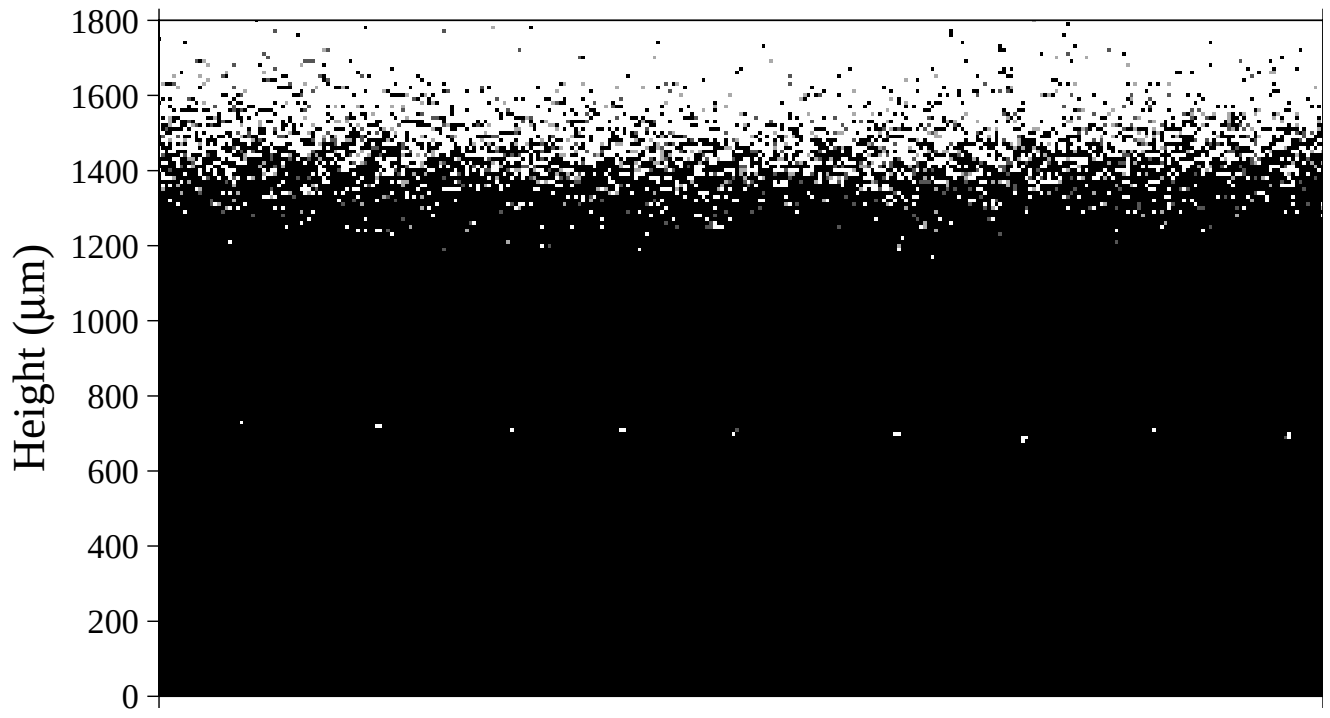

Fig. 3. Dust layer levitation height and the sheath edge. The pressure is 26.7 Pa, RF power is 5W, and DC bias is -68.5V, which is 19.7V below the natural bias of -48.8V.

In this case, Argon gas was employed at a pressure of 26.7 Pa (200 mTorr), the RF power is 5W, and the natural bias is

-48.8V. Figure 4 shows both the measured sheath thickness and the dust levitation height as a function of the difference in applied DC bias relative to their natural bias positions. The large uncertainty shown in the sheath thickness measurement is primarily due to the visual selection of the sheath edge, as shown in Fig. 3.

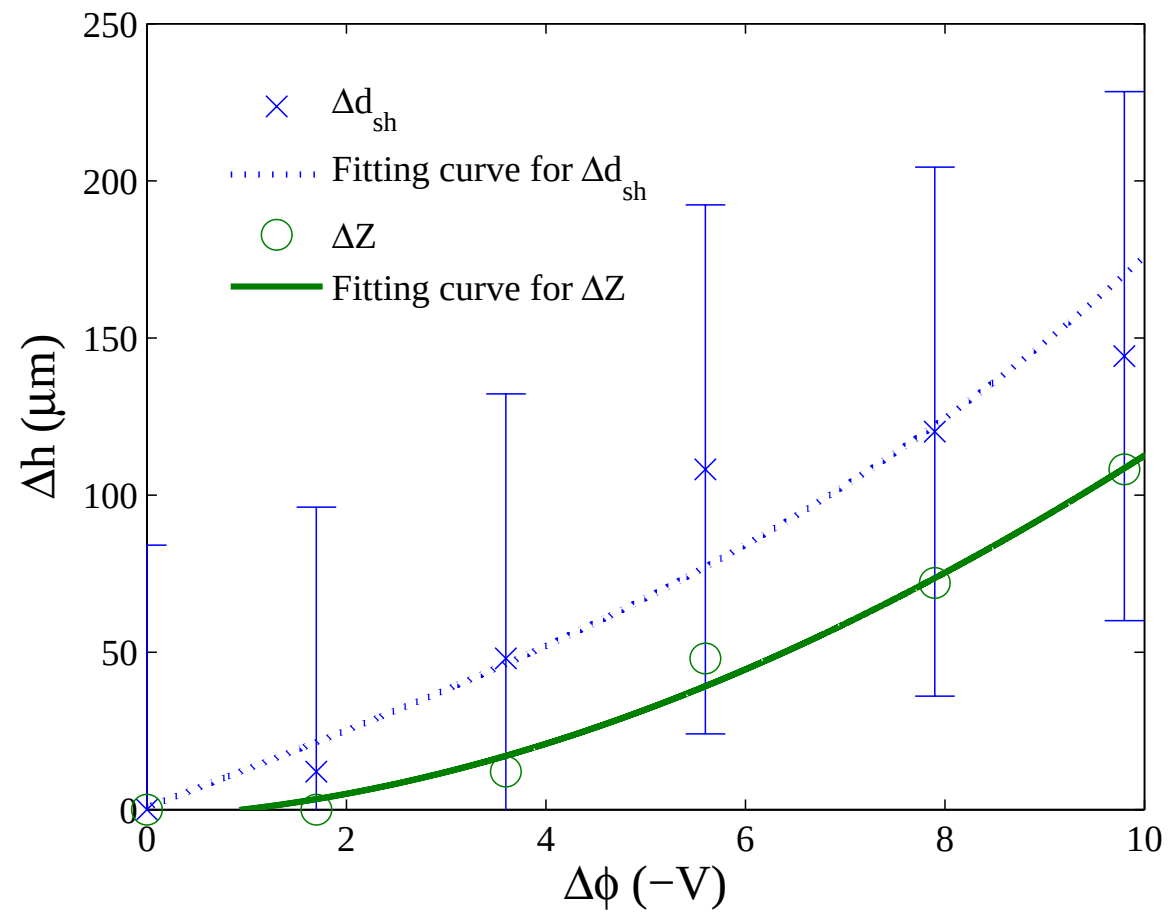

Fig. 4. Differences in dust layer height and sheath height as a function of the applied DC bias. As can be seen, as the DC bias difference $\Delta\phi \to 0$, the sheath thickness difference $\Delta d_{sh}$ and the dust layer difference $\Delta Z$ converge to zero.

As shown, as $\Delta\phi \to 0$, both $\Delta d_{sh}$ and $\Delta Z$ converge to zero, confirming $\Delta d_{sh} = \Delta Z$ for small changes in DC bias. Based on this, dust layer levitation height changes were used to represent the change in sheath thickness in the following calculations. Using the measured values of $\Delta Z$ and measured system parameters, the relationship of the Debye length as a function of the difference in DC bias $\Delta\phi$ was derived using Equations 3 and 4. The electron temperature used was measured to be $K_B T_e = 5.10eV$ (based on Langmuir probe data), and $\gamma = -0.5$ for the reasons described previously.

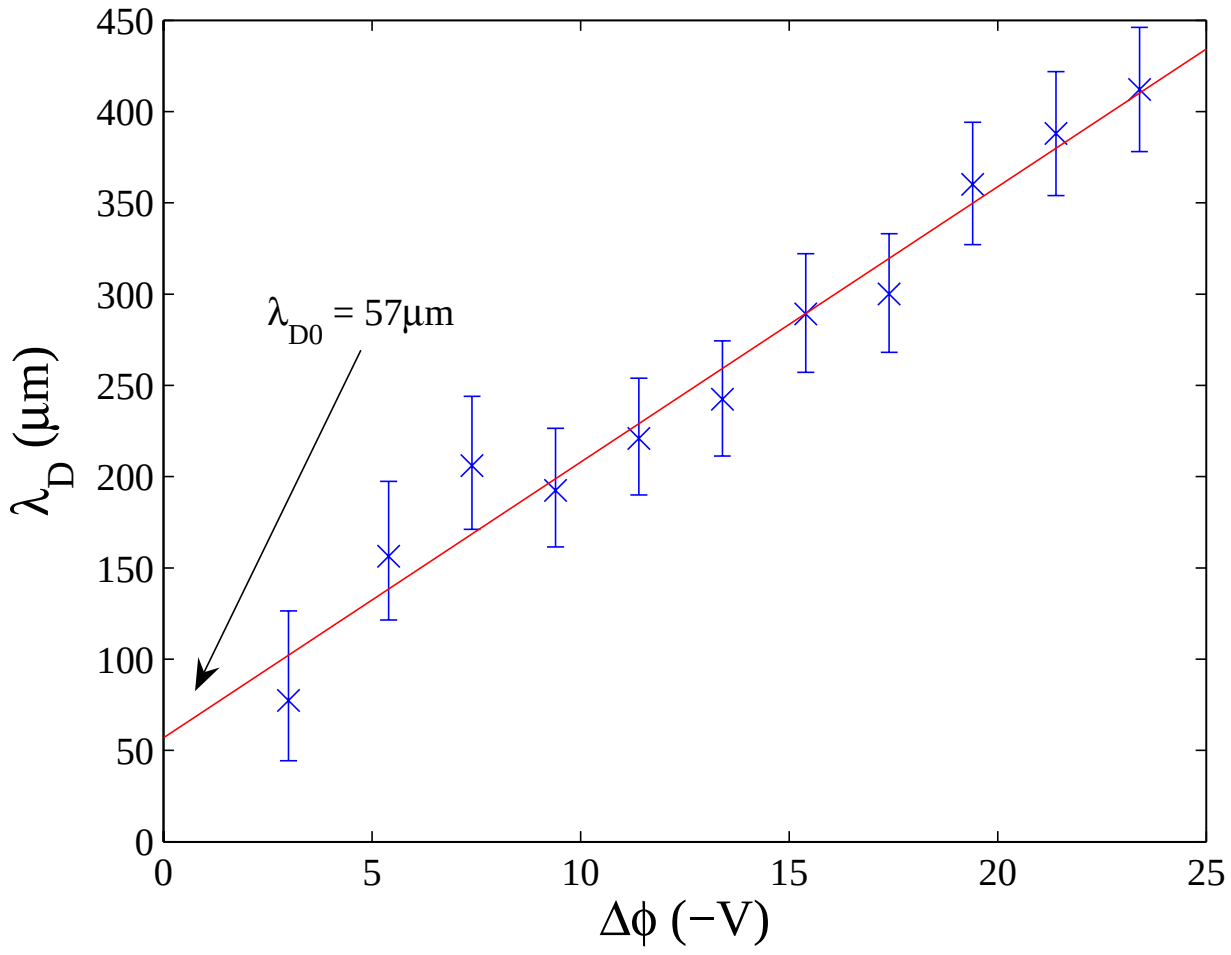

Fig. 5. The Debye length as a function of the DC bias difference, relative to a natural bias of -48.8V, derived by using Equations 3 and 4. The background gas is Argon, the pressure is 200 mTorr, and the rf power is 5W.

Using a best fit in Fig. 5, the Debye length was determined to be $\lambda_{D0} = 57 \pm 40\mu m$ at $\Delta\phi = 0$. As mentioned in the previous section, the error in this value includes the difference between $\gamma = 0$ and $\gamma = -1$, which accounts for about 25% of this uncertainty.

Employing the same technique, the Debye length $\lambda_{D0}$ at any DC bias other than the natural bias can also be determined experimentally. For example, Figure 6 shows the Debye length $\lambda_D$ as a function of $\Delta\phi$ around a bias voltage of -60.2V, yielding a value of $\lambda_{D0} = 234\mu m$.

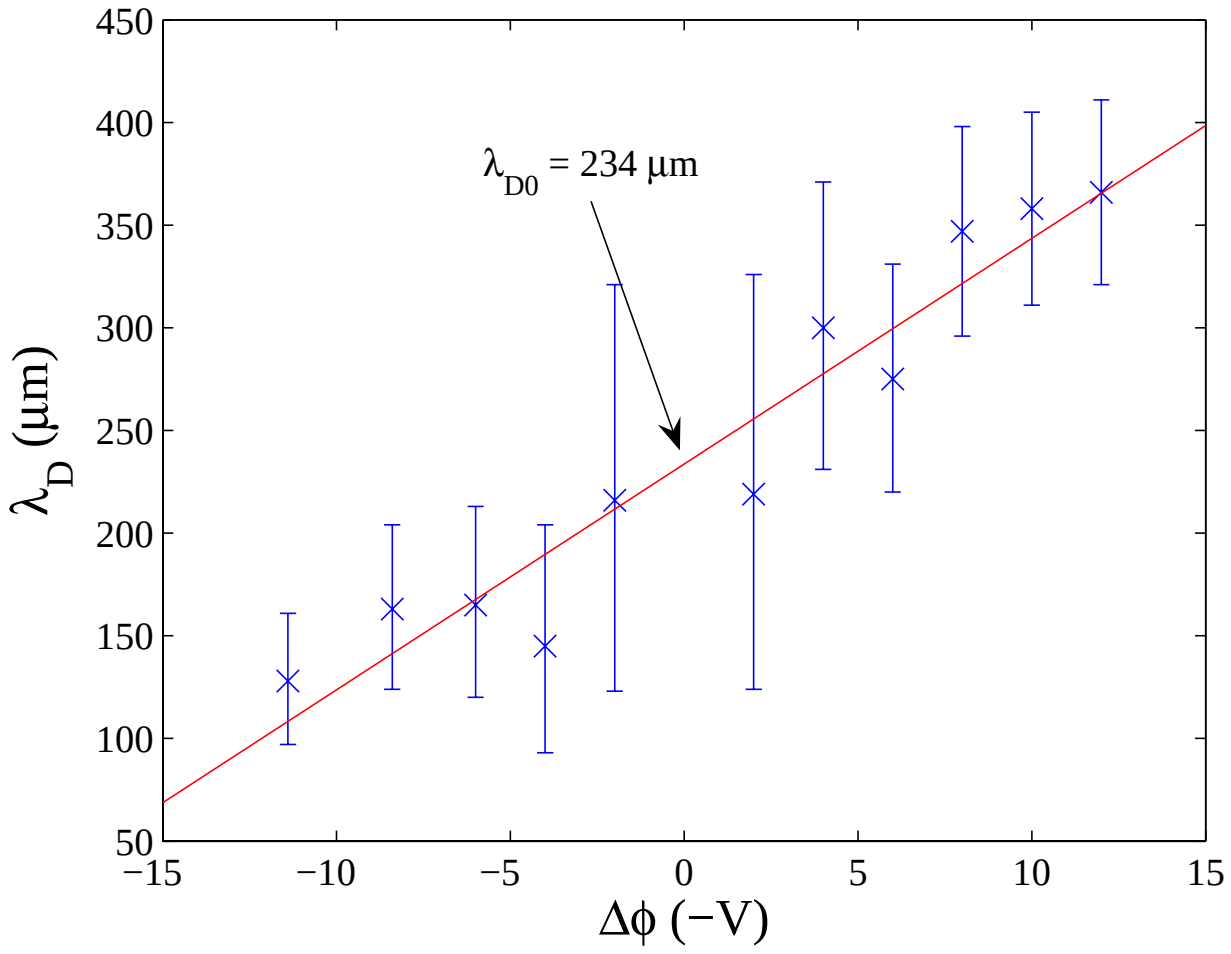

Fig. 6. The Debye length as a function of the DC bias difference, relative to an applied bias of -60.2V, as determined using Equations 3 and 4. The background gas is Argon, the pressure is 200 mTorr, and the rf power is 5W. As shown, $\lambda_{D0} = 234\mu m$ at $\Delta\phi = 0$.

Figure 7 shows $\lambda_{D0}$ as a function of the applied DC bias. From the fitting curve, the corrected Debye length under a natural bias, at an argon pressure of 200 mTorr, and a RF power of 5W for the CASPER GEC cell is $\lambda_{D0} = 79\mu m$.

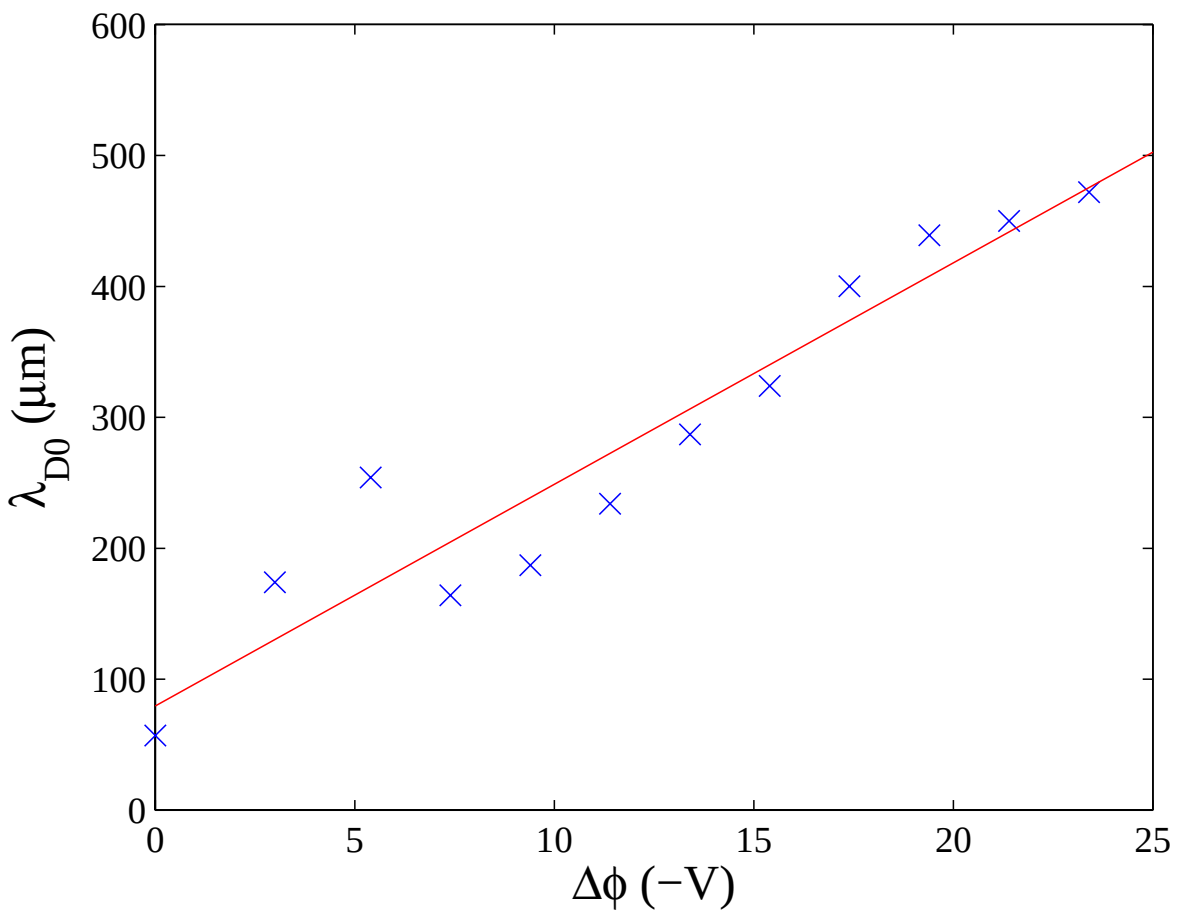

Fig. 7. Experimental results for the Debye length versus an applied DC bias. A general trend showing an increase in Debye length as the applied DC bias

increases negatively can be seen. The X-axis represents the difference in DC bias with respect to natural DC bias.

For comparison, Langmuir probe measurements were conducted using a SmartProbe from Scientific Systems Ltd. Under an argon pressure of 26.7 Pa (200 mTorr), a RF power of 5W, and allowing only natural bias, without dust particles, the ion density was measured to be $n_i = 5.71\times10^8 cm^{-3}$, the electron density was $n_e = 2.08\times10^8 cm^{-3}$, and the electron temperature was $T_e = 5.10eV$. The calculated Debye length for this case is $\lambda_D = 51.6\mu m$. Figure 8 shows the I-V curve for this Langmuir probe measurement.

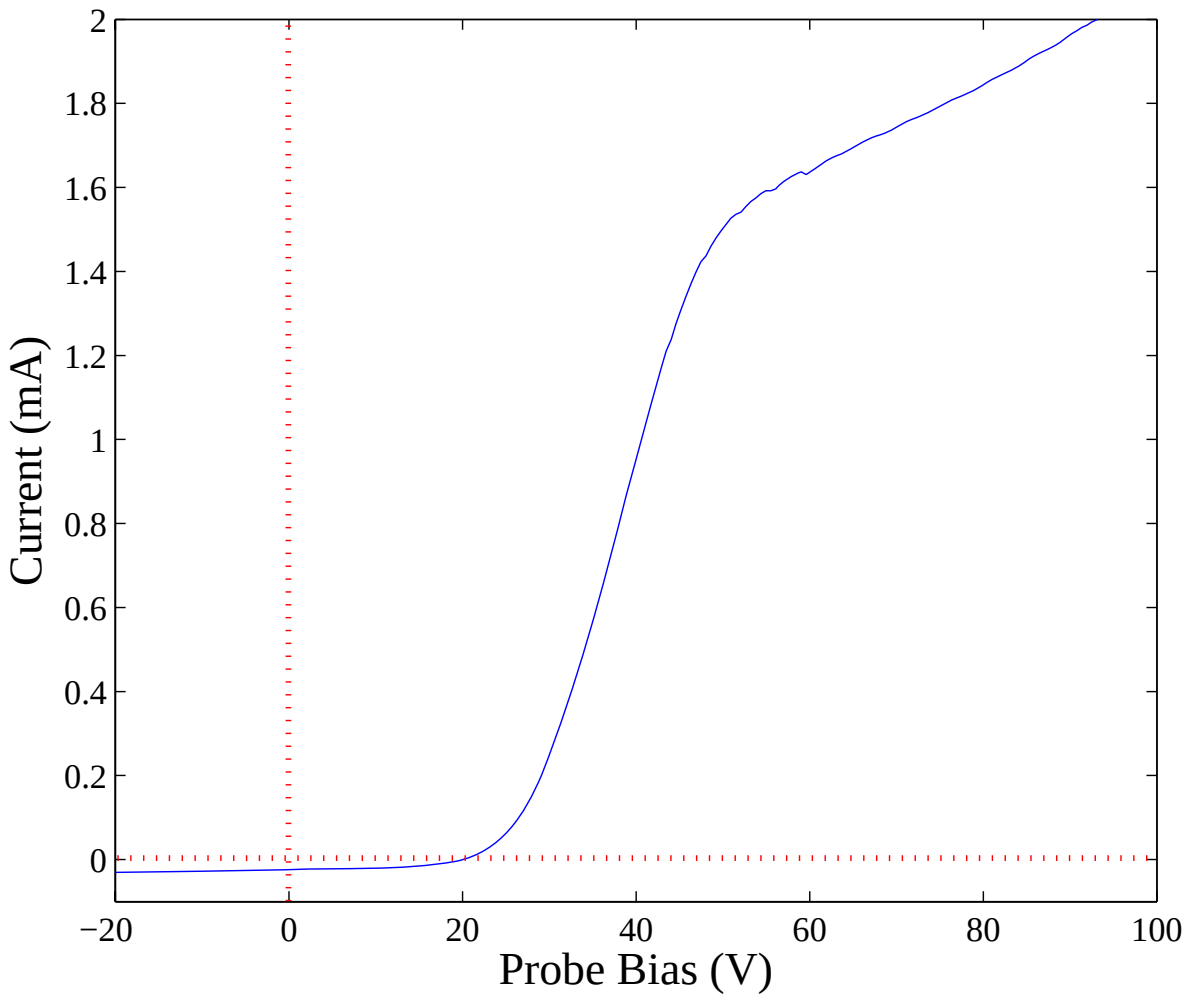

Fig. 8. I-V curve from representative Langmuir probe measurement. The argon pressure is 26.7 Pa, the RF power is 5W, and the system is under natural bias. The probe is positioned at the center of the reference cell directly above the lower electrode. The measured electron temperature was $T_e = 5.10eV$, the electron density was $n_e = 2.08\times10^8 cm^{-3}$, and the ion density was $n_i = 5.71\times10^8 cm^{-3}$.

## IV. DISCUSSION

There are several caveats to the above that should be discussed. First, employing the difference in sheath thickness (Equation 3) in all calculations rather than the direct thickness of the sheath (Equation 2) needs to be explained in greater detail. The primary reason for doing this is that using this approach reduces some of the systematic error. For example, Equation 2 is derived for a plasma without dust particles; however, recent simulation results show a difference between the dust sheath potential distribution for plasmas with and without dust [17]. As such, this difference in the potential distribution will introduce a small variance in the sheath thickness. Using the difference in the sheath thickness instead of the sheath thickness itself should minimize the error between the two cases.

Secondly, as shown in Figure 4, the difference between $\Delta d_{sh}$ and $\Delta Z$ increases as $\Delta\phi$ increases. One possible cause for this is the ion drag force. The ion drag force increases as the DC bias increases with the resultant ion streaming exerting an additional downward force on the dust particles due to the increased electric field created by the higher DC bias. This increases the error for larger values of $\Delta\phi$. Therefore, extrapolating between $\lambda_D$ and $\lambda_{D0}$, as in Figures 5 and 6, is another important and necessary step to minimize this error.

Finally, some of the parameters in Equation 4, such as $\gamma$ and $T_e$, must be determined employing other experimental techniques to insure an accurate determination of Debye length. These parameters can have a significant effect on the Debye length calculations. As mentioned above, the difference between $\gamma = 0$ and $\gamma = -1$ alone can result in a change in Debye length of about $\pm 40\mu m$ for a Debye length of $57\mu m$.

## CONCLUSIONS

A new experimental method for measuring the Debye length within a complex plasma has been introduced. This method is based on the change in dust levitation height caused by a change in the DC bias induced externally on the powered lower electrode. The result shows good agreement with Langmuir probe measurements. When compared to existing Debye length measurement techniques, the above method appears to be simpler in some cases to employ experimentally. It should be particularly suitable for small cluster Debye length measurements, since no external probe is necessary, minimizing perturbation problems.

**Truell W. Hyde** (Member) was born in Lubbock, TX in 1956. He received the B.S. degree in physics and math from Southern Nazarene University in 1978 and the Ph.D. degree in physics from Baylor University in 1988.
He is currently Professor of Physics, Director of the Center for Astrophysics, Space Physics & Engineering Research and the Vice Provost for Research at Baylor University.